\documentclass{article}
\usepackage{fixltx2e}
\usepackage{authblk}
\usepackage[latin1]{inputenc}
\usepackage{amsmath,amsfonts,amssymb,bm}
\usepackage{graphicx}
\usepackage{cite}
\usepackage[tight,hang,small,bf]{subfigure}

\begin{document}
\title{\textbf{A high-brightness source of polarization-entangled photons optimized for applications in free space}}
\author{Fabian Steinlechner,$^{1,*}$ Pavel Trojek,$^{2,3,4}$ Marc Jofre,$^1$\\ Henning Weier,$^{2,3}$ Daniel Perez,$^1$ Thomas Jennewein,$^5$ Rupert Ursin,$^6$ John Rarity,$^7$ Morgan W. Mitchell,$^{1,9}$ Juan P. Torres,$^{1,8}$\\ Harald Weinfurter,$^{3,4}$ and Valerio Pruneri$^{1,9}$}
\affil{\small{
$^1$ ICFO-Institut de Ciencies Fotoniques, 08860 Castelldefels (Barcelona), Spain \\
$^2$ qutools GmbH, 80539 M\"unchen, Germany \\
$^3$ Fakult\"at f\"ur Physik, Ludwig-Maximilians-Universit\"at M\"unchen, 80799 M\"unchen Germany \\
$^4$ Max-Planck-Institut f\"ur Quantenoptik, 85748 Garching, Germany \\ 
$^5$ Institute for Quantum Computing and Department of Physics and Astronomy, University of Waterloo, Ontario N2L 3G1, Canada \\ 
$^6$ Institute for Quantum Optics and Quantum Information, Austrian Academy of Sciences, Boltzmanngasse 3, 1090 Wien, Austria \\
$^7$ Department of Electrical and Electronic Engineering, University of Bristol, Bristol BS8 1UB, United Kingdom \\ 
$^8$ Department of Signal Theory and Communications, Universitat Politecnica de Catalunya, Jordi Girona 1-3, 08034 Barcelona Spain\\
$^{9}$ ICREA-Institució Catalana de Recerca i Estudis Avançats, 08010 Barcelona, Spain}}
\affil{*\underline{\emph{fabian.steinlechner@icfo.eu}}}
\date{}
\maketitle
\begin{abstract} We present a simple but highly efficient source of polarization-entangled photons based on spontaneous parametric down-conversion (SPDC) in bulk periodically poled potassium titanyl phosphate crystals (PPKTP) pumped by a 405 nm laser diode. Utilizing one of the highest available nonlinear coefficients in a non-degenerate, collinear type-0 phase-matching configuration, we generate polarization entanglement via the crossed-crystal scheme and detect 0.64 million photon pair events/s/mW, while maintaining an overlap fidelity with the ideal Bell state of 0.98 at a pump power of 0.025 mW.
\end{abstract}
\section{Introduction}
A number of key experiments \cite{Bouwmeester:97,Jennewein:00,Mattle:96} have shown that entangled photons are not only of paramount importance to questions regarding the nature of physical reality \cite{CHSH:69}, but can also find real-world applications in quantum communication. Considerable progress has been made in the distribution of entanglement over long-distance free-space links \cite{Ursin:07,Jin:2010}, with the next generation of envisaged projects likely to advance to experiments involving ground to satellite or inter-satellite links \cite{Rarity:02} ultimately enabling efficient quantum communication on a global scale. Many obstacles must be overcome towards the realization of world-wide quantum networks, in particular, the engineering of robust sources of entangled photons with sufficient brightness and entanglement visibility. Several schemes for the generation of photon pairs have been proposed and demonstrated \cite{Fulcois:07,Michler:02,Medic:10}, but at present by far the most convenient choice remains spontaneous parametric down-conversion (SPDC) in $\chi^{(2)}$ nonlinear crystals. In the first generation of type-II SPDC based sources \cite{Kwiat:95}, correlations in energy and momentum were exploited to generate polarization-entangled photons in restricted spatial regions. The pair collection efficiency was improved by overlapping type-I emission cones from separate crystals \cite{Kwiat:99}. The scheme was however limited to relatively short crystals due to the non-collinear configuration. A significant increase in the yield was obtained with a collinear non-degenerate configuration, which allowed the entangled photons to be efficiently coupled into single-mode fibers \cite{Trojek:08}. Further improvements were achieved in sources that made use of periodically poled materials \cite{Fedrizzi:07,Ljunggren:06,Kuklewicz:04}. Current developments in waveguide technology also present an attractive option for future entangled photon sources \cite{Martin:10,Fiorentino:07}.\\The approach presented in this paper is based on collinear, non-degenerate SPDC emission around 810 nm from two crossed type-0 PPKTP crystals pumped by a 405nm laser diode (LD), as previously demonstrated in \cite{Trojek:08}. With respect to this previous setup, which made use of beta-BaB$_2$O$_4$ crystals (BBO), we make use of two 20 mm periodically poled KTiOPO$_4$ (PPKTP) crystals. The crossed-crystal geometry allows accessing the largest nonlinear coefficient in PPKTP which is almost an order of magnitude larger than that utilized in other schemes and allows us to achieve unprecedented pair detection rates, under relaxed alignment requirements \cite{Fedrizzi:07,Hentschel:09,Kim:06}. High entanglement visibility was obtained by efficient cancellation of which-crystal information via an appropriately chosen Yttrium Vanadate (YVO$_4$) crystal \cite{Ljunggren:06,Trojek:Phd,Nambu02}. The tailored nonlinearity applied here enables the use of well-developed blue laser diodes to generate photon pairs in a spectral region efficiently detectable with silicon avalanche photodiodes \cite{Gallivanoni:06}. The simple, inherently robust set-up, together with the high yield of photon pairs, make our source of entangled photon pairs a valuable tool for the envisaged future field trials over atmospheric free-space links.

\section{Type-0 SPDC in PPKTP}
\begin{figure}[htbp]
\centering\includegraphics[width=8cm]{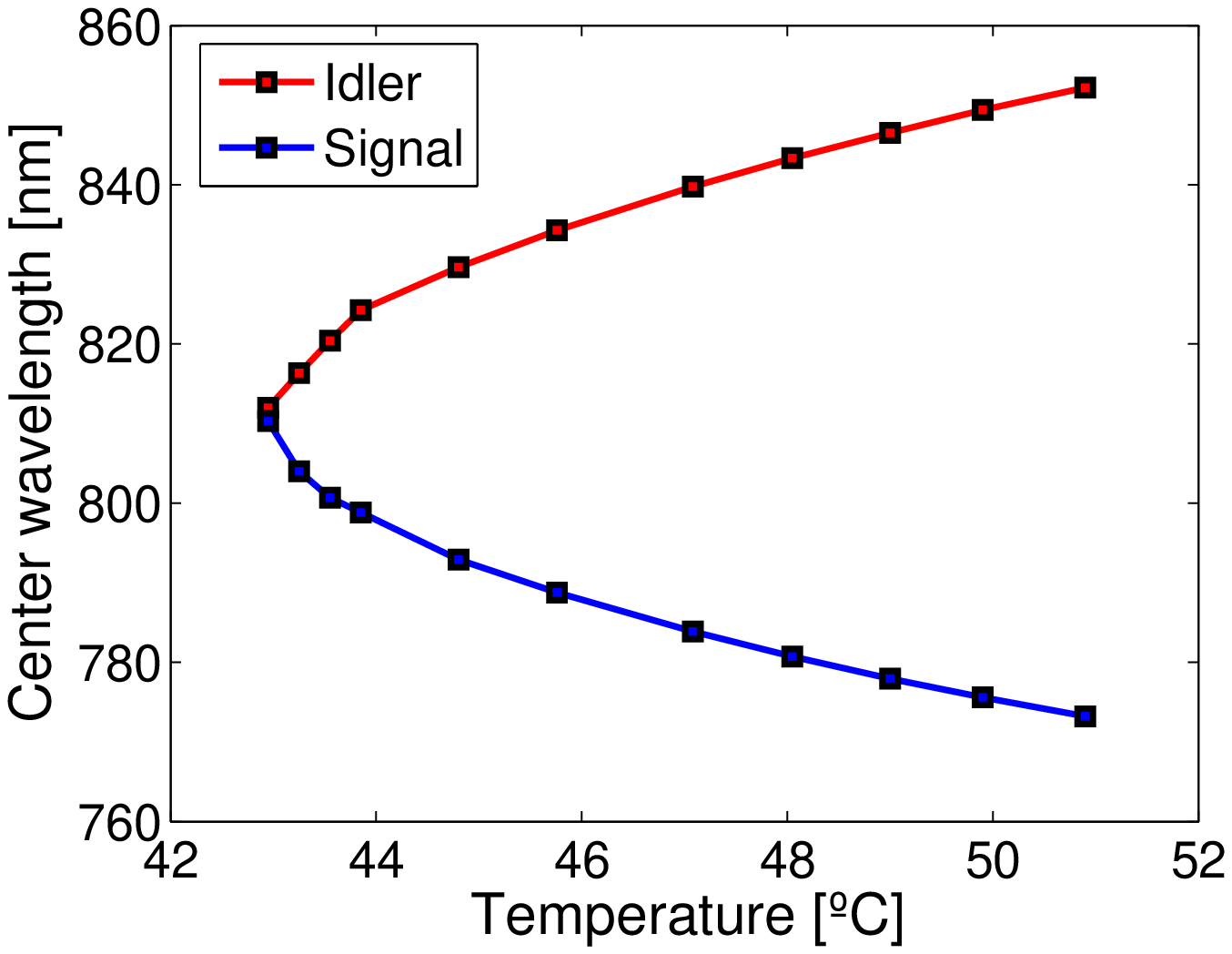}
\centering\includegraphics[width=8cm]{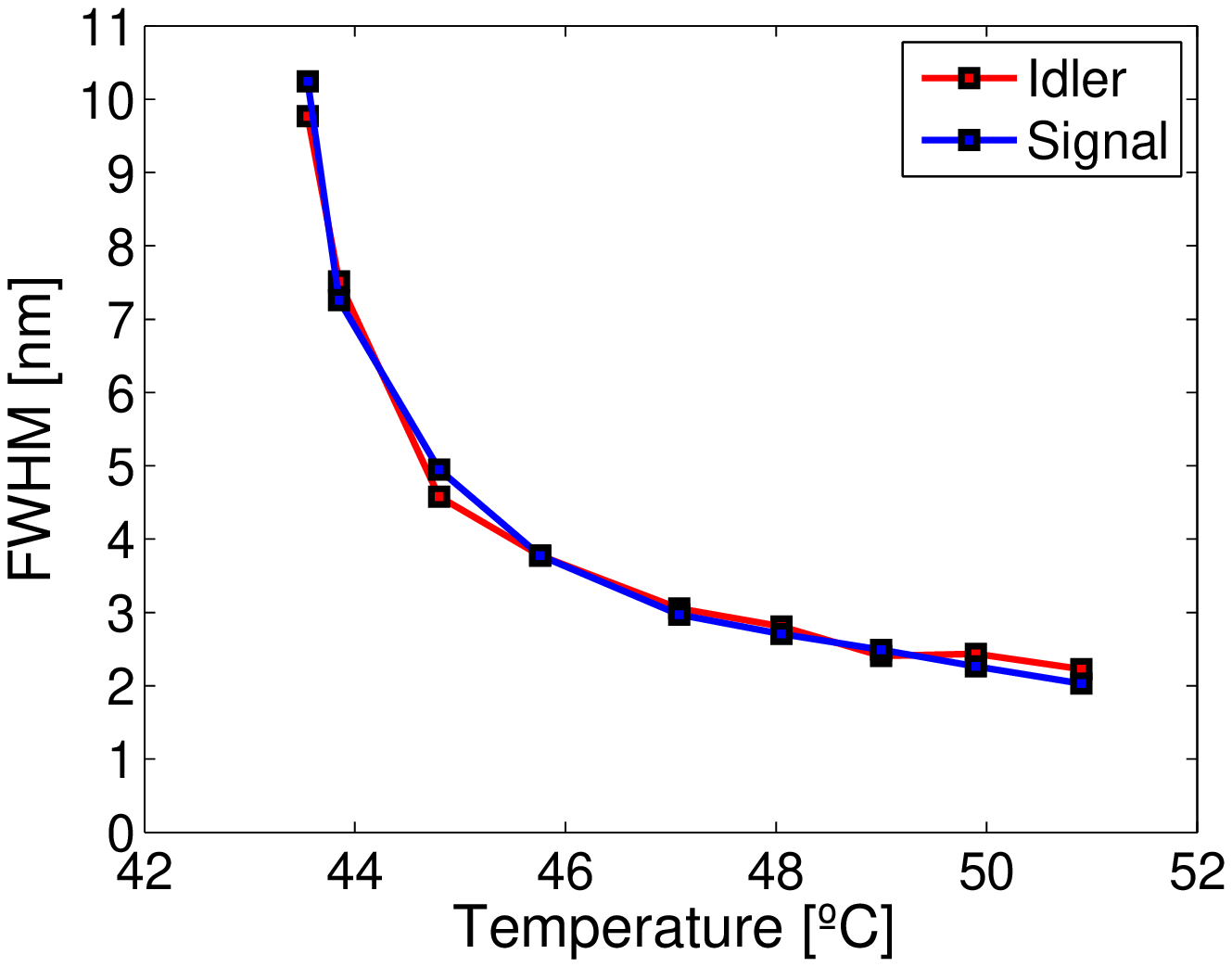}
\caption{Measured phase-matching characteristics from experimental data. (left) Phase-matching curve showing increased non-degeneracy of signal (blue) and idler (red) photons as temperature increased above SHG point. (right) FWHM bandwidth of signal and idler photons as a function of temperature.}
\label{SPCDCWLFWHM}
\end{figure}

Spontaneous parametric down-conversion in second-order nonlinear crystals is a process in which atoms of the crystal mediate the conversion of a high energy pump photon into two photons with lower frequency, conventionally termed signal and idler photons. Depending on the polarizations of the interacting fields, one distinguishes between type-0, type-I and type-II SPDC, whereby type-I and II processes involve orthogonally polarized fields, whereas in type-0 processes all photons are co-polarized.  
\begin{figure}[htbp]
\centering\includegraphics[width=12cm]{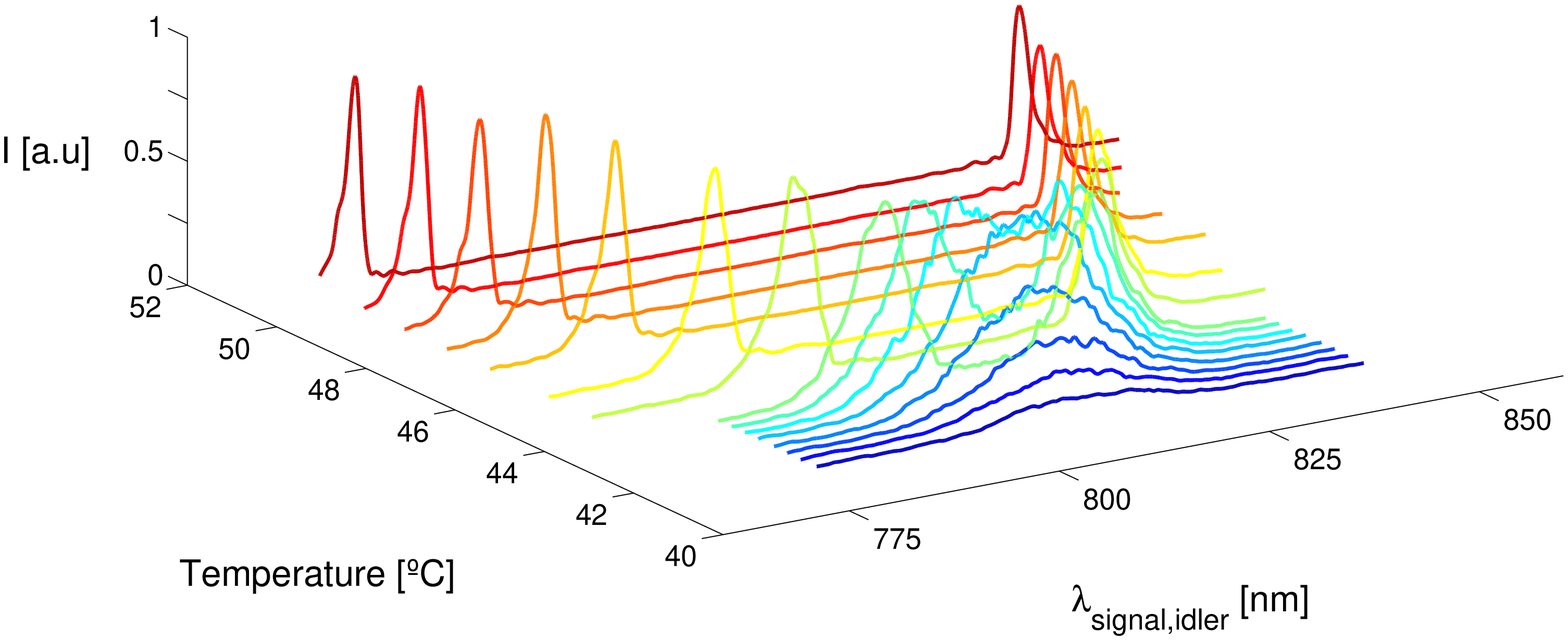}
\caption{Experimentally observed normalized SPDC spectra for 20 mm PPKTP crystal (poling period=$3.425\mu$m) pumped with a 405.4 nm CW laser diode for varied phase-matching temperature. The photons were coupled into a single-mode fiber and analyzed in a grating spectrometer with a resolution of 0.4 nm}
\label{SPDC}
\end{figure}
SPDC is possible for wavelengths that fulfill both energy conservation $\omega_p=\omega_s+\omega_i$ as well a phase-matching conditions $\Delta \vec{k} = \vec{k_p} - \vec{k_s} - \vec{k_i} - \vec{\frac{2\pi}{\Lambda}}$ (here $\omega_{p,s,i}$ denote the angular frequencies of the respective pump, signal and idler photons, while $\vec{k}_{p,s,i}$ denote the respective wave-vectors). Recent advances in periodic poling technology, achieving increasingly shorter poling periods $\Lambda$, allow first-order quasi-phase-matching for almost any wavelengths and polarizations. The spectral properties of the SPDC emission are given by 
\begin{equation}
I(\lambda)\propto sinc^2(\frac{\Delta k L}{2})
\end{equation}
resulting in an SPDC full-width half-maximum (FWHM) bandwidth, which, for sufficiently non-degenerate signal and idler wavelengths, is inversely proportional to the crystal length $L$. Figure \ref{SPDC} shows the temperature-dependent SPDC response of photon pairs emitted collinearly from a 20 mm flux grown type-0 PPKTP crystal \cite{Satyanarayan:99} with a poling period of 3.425  $\mu$m, pumped by a single-frequency laser diode with a center wavelength (CWL) of 405.4 nm. From this we determined the SPDC CWL and FWHM of signal and idler photons shown in Fig. \ref{SPCDCWLFWHM}.
An important parameter subject to optimization when designing an efficient entangled photon source, is the focus geometry. It has been shown in several theoretical and experimental studies \cite{Fedrizzi:07,Ljunggren:05,Palacios:11}, that efficient coupling of SPDC into single-mode fibers (SMF) is achieved by selecting the confocal beam parameters of pump, signal and idler fields appropriately. We performed an experimental optimization of these parameters for our 20 mm type-0 PPKTP crystals and found the highest coincidence rates for a waist combination of $w_p\sim18 \mu$m and $w_{s,i}\sim24 \mu$m for the pump, signal and idler modes, respectively. With the pump waist fixed to its optimal value, we see that the maximal coincidences were obtained by setting $w_{s,i}\sim\sqrt{2} w_{p}$. This can be understood as matching the Rayleigh ranges $z_R=w^2\pi/\lambda$ of SPDC and pump fields to optimize coupling. Note that these values are consistent with the optimal values of $w_p\sim22 \mu m$ and $w_s\sim31 \mu m$ predicted in \cite{Palacios:11}.

\section{Entangled photon source}
\begin{figure}[htbp]
\centering\includegraphics[width=8cm]{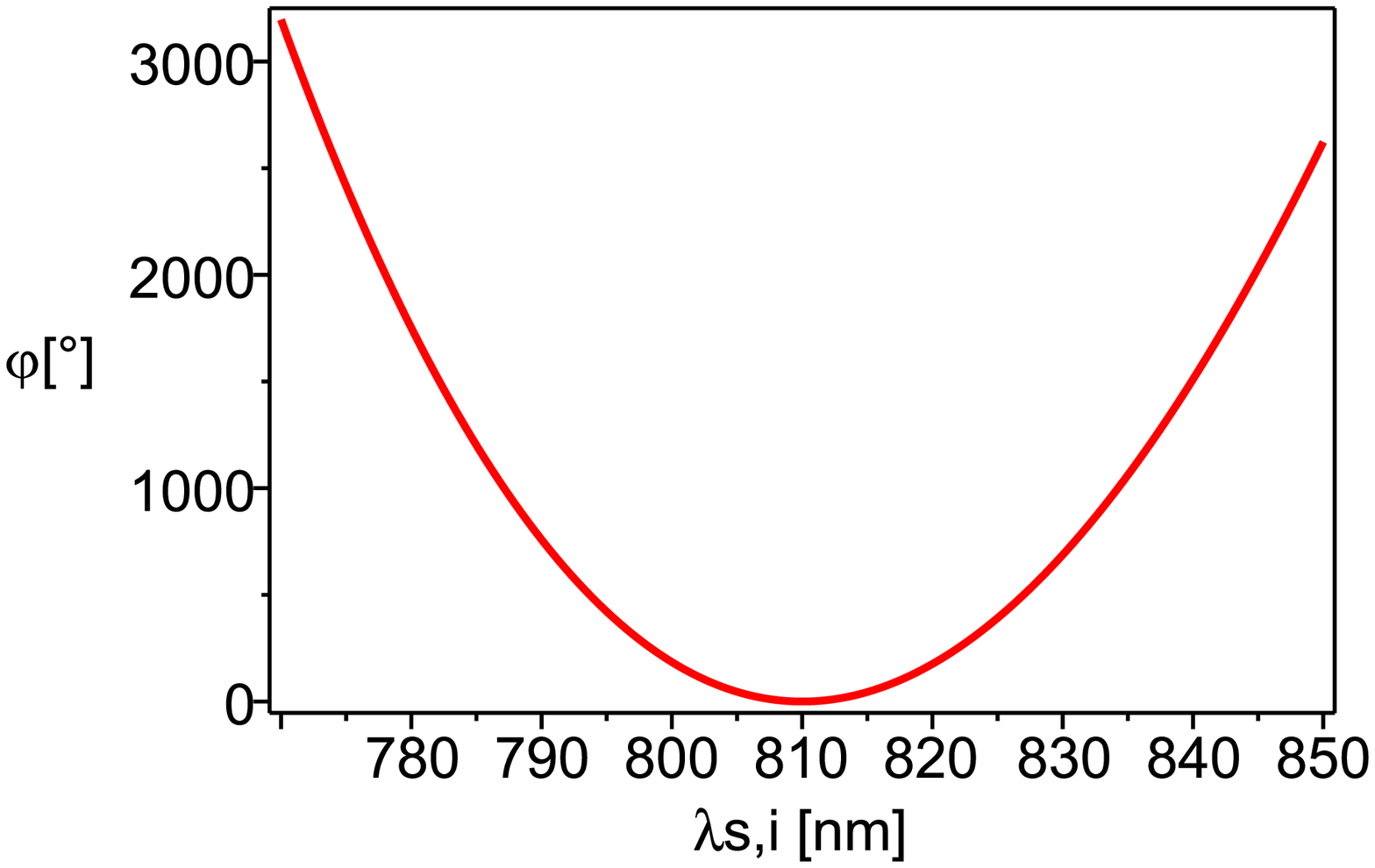}
\centering\includegraphics[width=8cm]{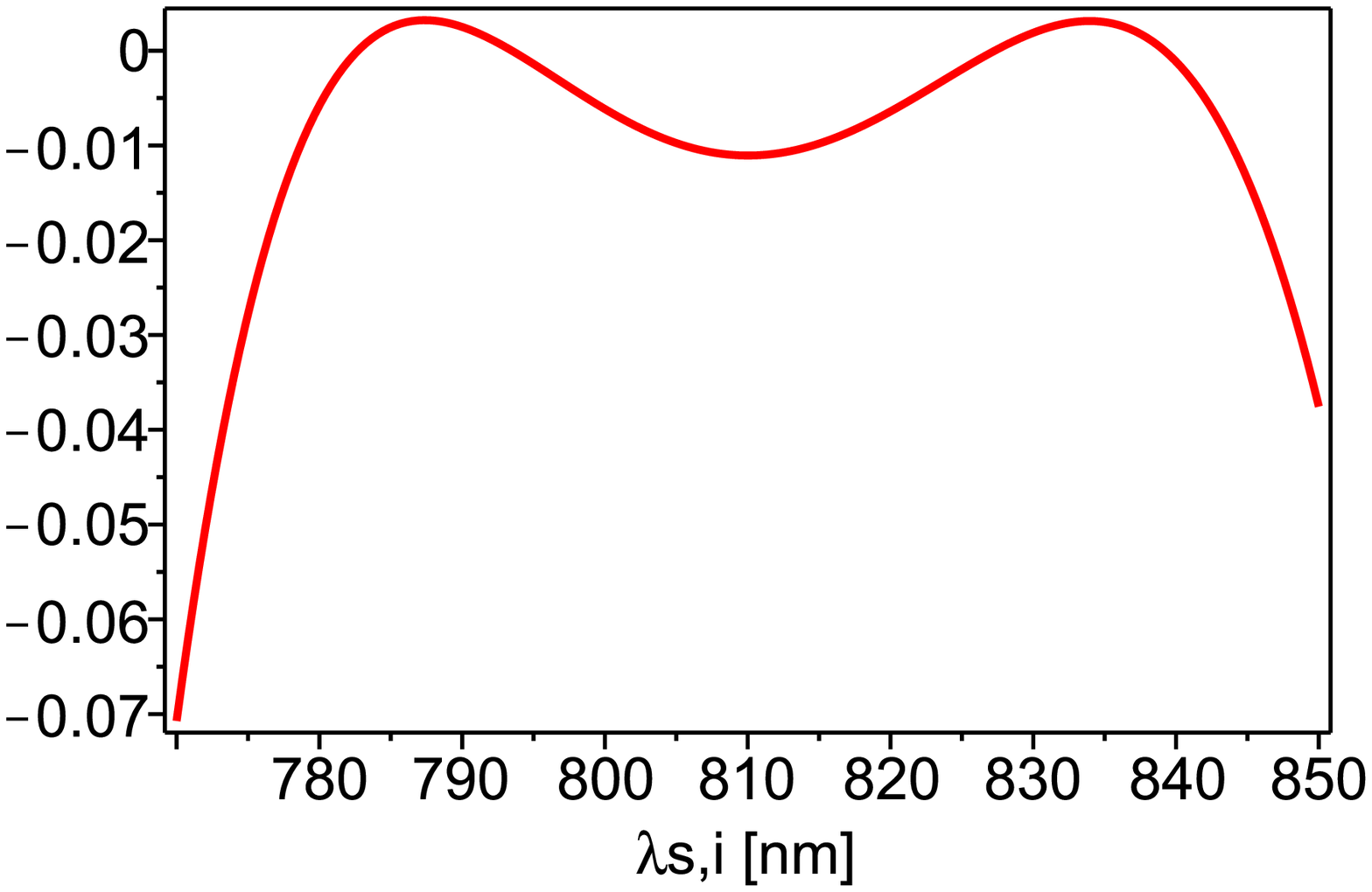}
\caption{(left) Strongly varying phase dependence for 2x20 mm PPKTP due to additional chromatic dispersion experienced by photon pairs emitted in first crystal. (right) Flattened dependence after compensation with 30 mm YVO$_4$ crystal, with negligible variation over broad spectral range; note the difference in scale.}
\label{phasemap}
\end{figure}

We generate co-polarized photon pairs at non-degenerate wavelength around 810 nm via the large nonlinearity $d_{eff}=\frac{2}{\pi}d_{33}=$ 12 pm/V of KTP \cite{Satyanarayan:99}. This operating wavelength was selected with respect to possible free-space communication applications, as an optimized trade-off between atmospheric transmission, beam divergence over long-distance links, detector efficiency and the availability of compact LD pump sources. The periodic poling of the crystal allows tailoring the interaction to a collinear non-critical phase-matching condition. This eliminates spatial walk-off and allows efficient coupling into a single-mode fiber, where the signal and idler photons are spectrally isolated using a fiber-based wavelength division multiplexer (WDM). The single-mode WDM output provides a versatile plug-and-play interface and ensures minimum diffraction over long distance links. Polarization entanglement is generated in a crossed-crystal configuration \cite{Kwiat:99}. In this scheme two mutually orthogonally oriented PPKTP crystals are placed directly one after the other. This double-crystal configuration is pumped with UV-light polarized diagonally with respect to the two crystallographic z-axis, which leads to equal probabilities for photon pair emission from the first crystal (in state $\vert H_s H_i \rangle$) or from the second one (state $\vert V_s V_i \rangle$). Thus a polarization-entangled state 
\begin{equation}
\vert \Psi \rangle = \vert V_s V_i \rangle+ e^{i \phi(\lambda_s,\lambda_i)} \vert H_s H_i \rangle
\end{equation}
is generated, if a fixed relative phase relationship $\phi(\lambda_s,\lambda_i)$ is maintained over the detected SPDC bandwidth. We note that any spatial which-crystal information the photons may have carried after SPDC emission is effectively erased via the projection into the single spatial mode of the coupling fiber. The purity of polarization entanglement is thus not impaired by any spatial effect. \\However, as a consequence of the crossed-crystal geometry, the  $\vert H_s H_i \rangle$  pairs generated in the first crystal acquire an additional phase shift relative to the  $\vert V_s V_i \rangle$ pairs emitted from the second. As this can yield which-crystal information that leads to de-phasing of the entangled state, compensation of this phase is required. To briefly discuss the compensation scheme, we compare photon pairs generated at the input facet of the first and second crystal. In this case the accumulated phase difference, resulting from individually acquired phases $\phi^{H,V}_{s,i}$ of horizontally and vertically polarized signal and idler photons at the output of the second crystal, reads  
\begin{equation}
\phi(\lambda_s,\lambda_i) =\phi_p+\phi^H_s+\phi^H_i-(\phi^V_s+\phi^V_i) = \phi_p + 2 \pi L \left(\frac{n_y(\lambda_s)}{\lambda_s}+\frac{n_y(\lambda_i)}{\lambda_i}\right)
\end{equation}
whereby any $n_z$ terms cancel due to the phase-matching condition in the two crystals. In the case of 2x20 mm PPKTP crystals, this leads to strong variations of the relative phase depicted in Fig. \ref{phasemap}. In order to counteract this de-phasing effect, a 30.01 mm YVO$_4$ crystal exhibiting opposite birefringent characteristics $\phi_{C}(\lambda_s,\lambda_i)$ is inserted after the down-conversion crystals. According to
\begin{equation}
\phi_{C}(\lambda_s,\lambda_i) = 2 \pi \times L_{YVO}\left[\frac{n^{(o)}(\lambda_s)}{\lambda_s}+\frac{n^{(o)}(\lambda_i)}{\lambda_i}-\left( \frac{n^{(e)}(\lambda_s)}{\lambda_s}+\frac{n^{(e)}(\lambda_i)}{\lambda_i} \right) \right]
\end{equation}
it becomes possible to effectively flatten the phase dependence over a broad spectral range. Note that we neglect de-phasing effects present in the case of a broad pump bandwidth. This is, however, not a limitation, since such effects can be pre-compensated, as demonstrated in \cite{Trojek:08,Nambu02}, still allowing the use of free-running or pulsed pump sources. A more detailed calculation in the time domain can be found in \cite{Trojek:Phd,Nambu02}. 
\section{Experimental setup}
\begin{figure}[htbp]
\centering\includegraphics[width=10cm]{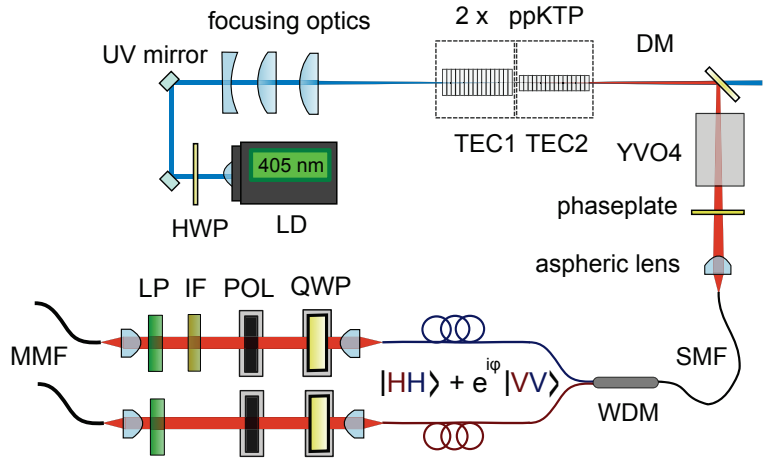}
\caption{Schematic experimental setup. The pump LD is corrected for astigmatism and focused to the center interface of two thermally controlled (TEC1,TEC2) PPKTP crystals. The SPDC generated in the crystals is separated from the pump photons via a dichroic mirror (DM) and passes a YVO compensation crystal and phase-plate, before being coupled into a SMF that guides the photons to the output ports of a WDM. The source including driving electronics are mounted on a compact 40 x 40 cm$^2$ breadboard. The WDM output fibers are connected to free-space polarization analyzers, consisting of a quarter wave plate (QWP) and a polarizer (POL), where further filtering occurs via an interference filter (IF) and longpass filters (LP). The photons are then detected via multi-mode fiber (MMF) coupled silicon SPADs.}
\label{schematicsetup}
\end{figure}

In the experimental setup (Fig. \ref{schematicsetup}), the astigmatic output beam of a volume holographic stabilized laser diode (Ondax Inc.) with a center wavelength of 405.04 nm is corrected by a 2:1 astigmatism telescope consisting of two cylindrical lenses. The circular pump beam is then focused to the optimized waist size of approximately 18 $\mu$m \cite{Ljunggren:05,Palacios:11} at the center interface of two 20 mm crossed PPKTP crystals. The crystals have a poling period of 3.425 $\mu$m for type-0 collinear phase-matching from 405 nm to the non-degenerate wavelengths 783nm (signal) and 837 nm (idler). The two crystals are mounted on a double-oven consisting of two individual Peltier elements to account for differences in the crystals and maintain them at phase-matching temperatures for equal center wavelengths. For our crystals this amounted to a temperature difference of 0.22°C (most likely due to imperfections in the crystal growing process resulting in inhomogeneities of the refractive index). The SPDC photons inherit a mode with a waist size of $\sim$ 24 $\mu$m and propagate collinearly with the pump laser. A dichroic mirror (DM) transmits 99\% of the pump photons and reflects the SPDC photons through a 30.01 mm YVO$_4$ crystal, which reverses dispersive de-phasing effects. An additional 100 $\mu$m thick YVO$_4$ plate was angle tuned to set the relative phase of the polarization-entangled state. A color-glass long pass filter isolates the remaining pump photons and rejects stray light. Using a f =11 mm aspheric lens, the SPDC photons are coupled into a single-mode fiber, guiding the photons to a WDM which splits signal and idler photons with non-degenerate wavelengths into two output fibers. To assess the performance of the source, the output ports of the WDM were each collimated and sent to a free-space single-port polarization analyzer, consisting of a quarter-wave plate (QWP) and a thin-film polarizer. The photons are then further filtered via an interference filter with a FWHM of 3.5 nm and a peak transmission of $\sim$90\%, placed in signal path. This surpessed broadband background emission from the PPKTP crystals and increased the spectral overlap of SPDC from the two crystals (see Fig.\ref{spectra}). After traversing the analyzer modules, the signal and idler photons are coupled into multi-mode fibers and guided to two single-photon avalance diodes (SPAD) with an approximate detection efficiency of $\sim40\%$ ($\sim$500 cps dark counts), where coincident measurement events were recorded via a fast time-to-digital converter (quTAU) with the coincidence window set to 2.4 ns.

\section{Results}
\begin{figure}[htbp]
\centering\includegraphics[width=6.5cm]{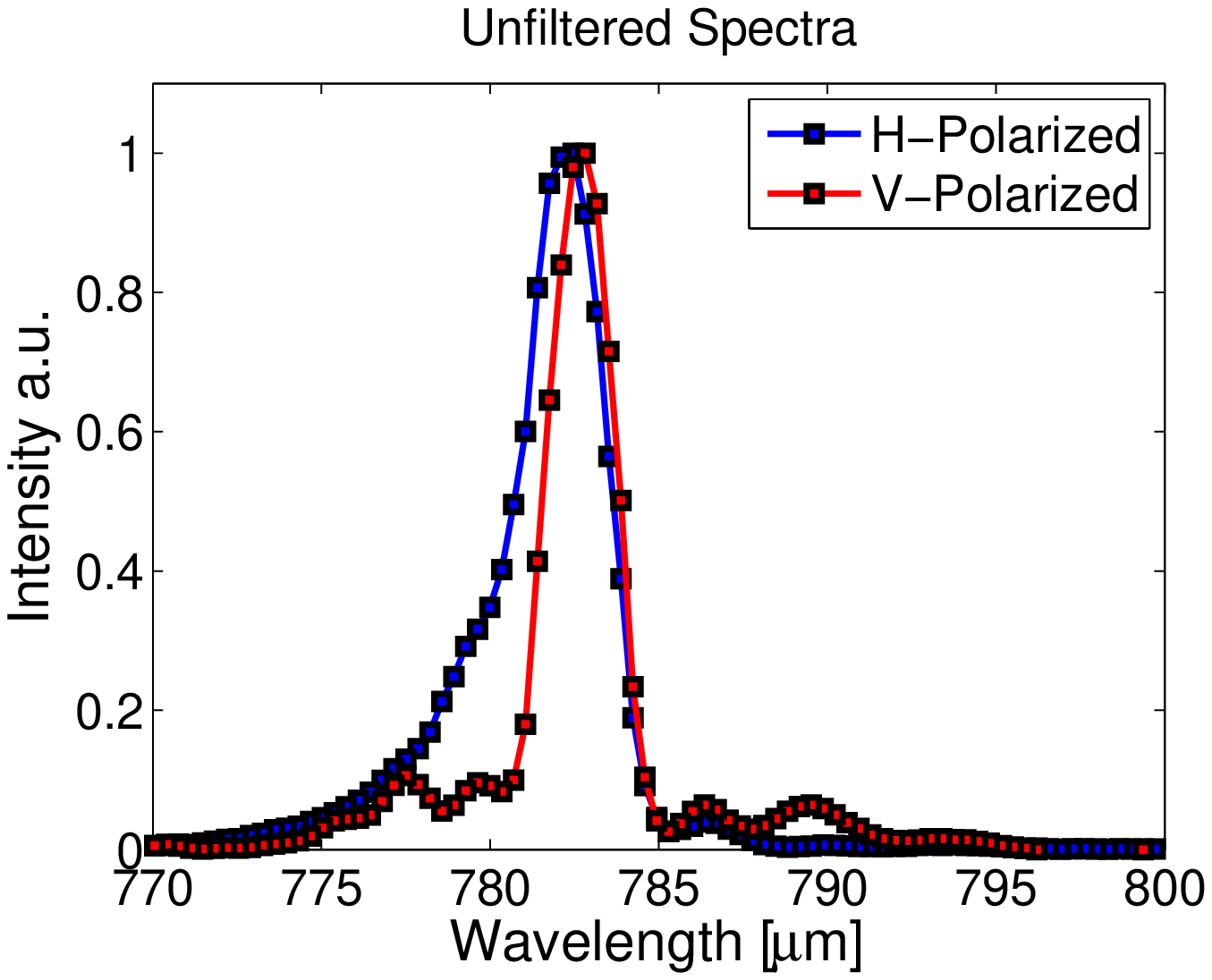}\centering\includegraphics[width=6.5cm]{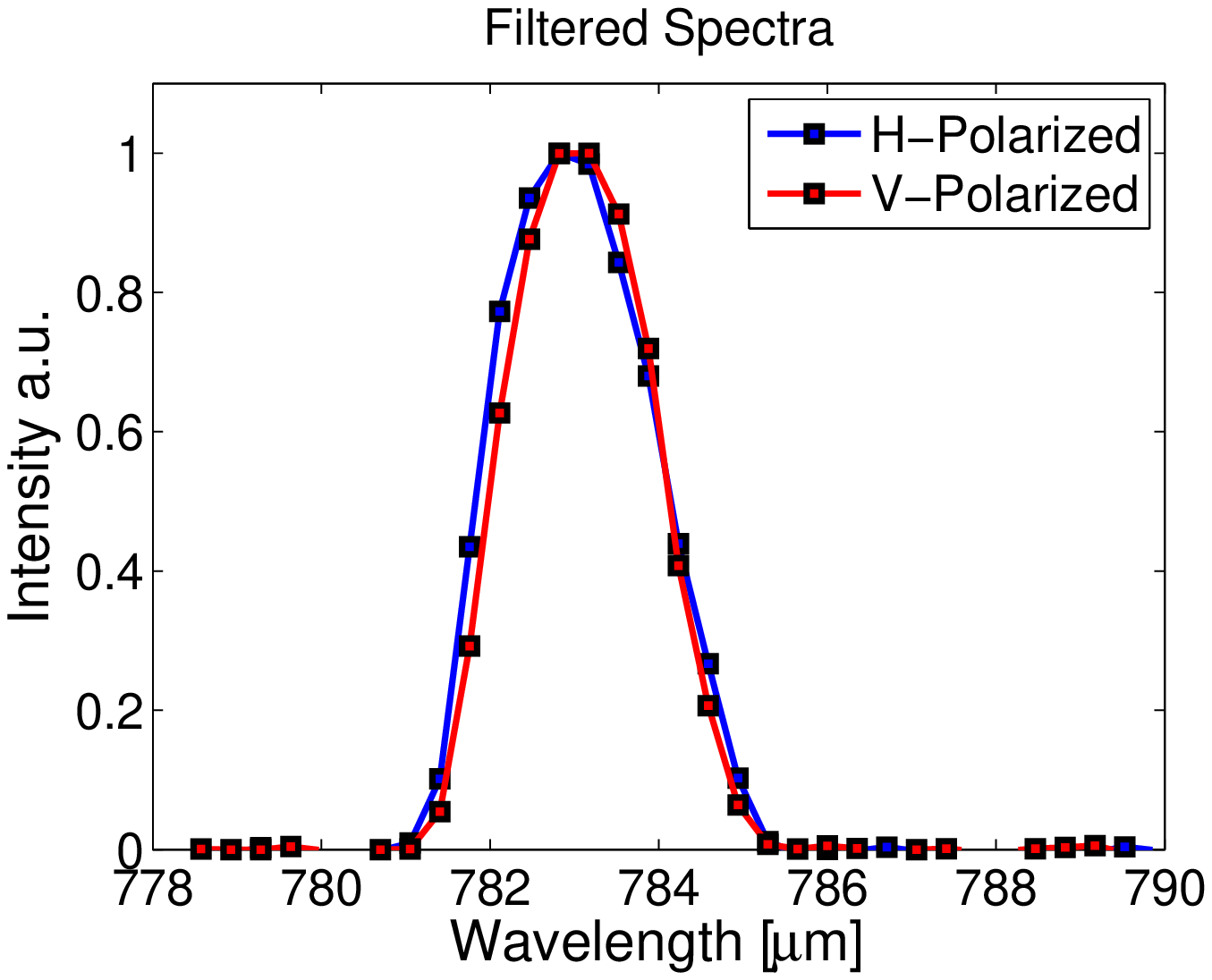}
\caption{SPDC spectra of horizontally and vertically polarized signal photons before and after filtering with a 3.5 nm FWHM IF filter. The spectra were recorded with the crystals maintained at phase-matching temperatures of 28.3ºC (H-Polarization) and 28.1ºC (V-Polarization). The spectral intensity overlap integrals were calculated to 91 \% and $>$99\% for the unfiltered and filtered SPDC profiles, respectively.}
\label{spectra}
\end{figure}

\begin{figure}[htbp]
\centering\includegraphics[width=13cm]{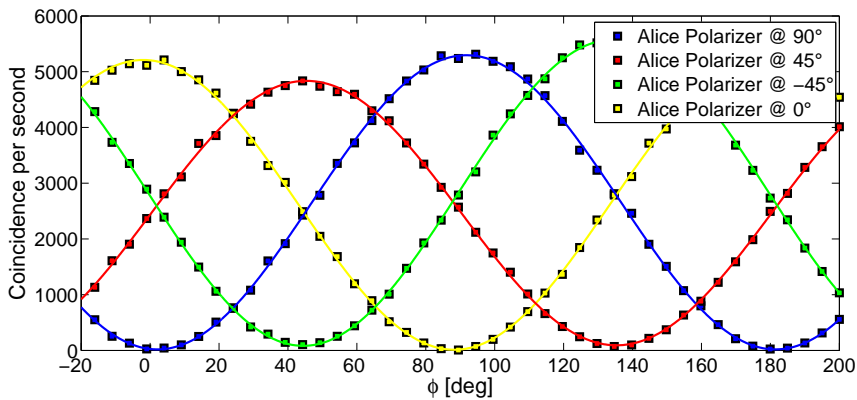}
\caption{Correlation functions obtained for diagonal D/A (red/green) and H/V (yellow/blue) measurement basis. Pump power 0.025 mW. (Square experimental data-points, line : best fit). The visibilities calculated from the fits to raw data (no correction for accidental coincidence counts) were H/V 99.5\%/99.3\%, D/A 97.0\%/96.2\%}
\label{fig:correlations}
\end{figure}

With the polarizers removed from the polarization analyzer, but the spectral filters in place, we detect a total coincidence rate of $R_c$ = 16000 cps and a signal singles rate $R_s$ =89000 cps at a pump power of merely 0.025 mW. Together with a measured FWHM of 2.3 nm, these values amount to a \emph{detected} pair rate of 640 kcps/mW and a detected spectral brightness of 278 kcps/mW/nm at a conditional coincidence ratio $\frac{R_c}{R_s}=$ 0.18: to our knowledge the highest reported normalized detected pair rate for this type of system.\\The polarization entanglement was characterized by measuring the  polarization correlation functions in two mutually unbiased bases (Fig. \ref{fig:correlations}). To more completely assess the degree of entanglement a quantum state tomography was performed, whereby a Bell state fidelity $F= \langle \Phi^+ \vert  \rho  \vert \Phi^+ \rangle$ of 0.983 $\pm$ 0.005 was achieved at a pump power of 0.025 mW. As discussed in the following, accidental coincidences were negligible at such low pump powers. We attribute the bounded visibility to remaining which-crystal information due to non-identical SPDC spectra of the two crystals (overlap integral $\sim$99\%, see Fig. \ref{spectra}) and imperfect timing compensation, as well as a weak polarization dependence of the WDM splitting ratio ($\sim 10\%$ variation with input polarization). Scaling from our detected brightness at low pump powers we estimate a detected coincidence rate larger than 20 Mcps locally detectable at 40 mW pump power. With the typical saturation rate of 10 Mcps for commercial silicon SPADs, the total singles count rates of $2\times R_c/\eta_c\sim$ 200 Mcps would already require an array of about 60 detectors for registration of the pair coincidences at $\sim$30\% of the detector saturation level. In addition to this immense logistic requirement there is a fundamental limitation, due to multi-pair emission in SPDC.

\section{Multi-pair limited entanglement visibility at high pump power}
\begin{figure}[htbp]
\centering\includegraphics[width=13cm]{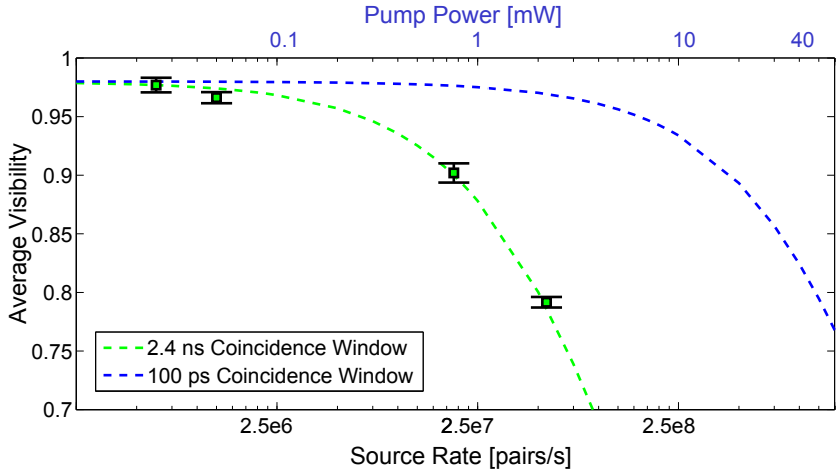}
\caption{Polarization correlation visibility obtained from full tomographic density matrix reconstructions at 0.025 mW, 0.05 mW, 0.78  mW and 2.2  mW respectively. The measurements at 0.78 mW and 2.2 mW were conducted with attenuation added to both paths of the polarization analyzer (Fig. \ref{schematicsetup}) to avoid SPAD saturation. The bottom scale denotes the estimated number of photon pairs at the crystal output. We evaluate the impact of multi-pair emissions via a Matlab simulation of the SPDC photon pair number, utilizing a quantum optics toolbox \cite{JenneweinToolbox}. The green line indicates the modeled behavior for pair generation rates scaled to higher pump powers. The visibility of the emitted state was set to 98\% to account for other experimental imperfections, as outlined in the previous section. Even for a coincidence window of 100 ps (dotted blue line) the simulation shows that a visibility above 90 \% is no longer maintainable for  pump powers above 20 mW.}
\label{multipairs}
\end{figure}

At high pump powers, we observe a drastic reduction of entanglement visibility. This reduction of visibility at increased source rates is due to the random pair emission characteristics of SPDC, in which there is always a finite probability of emitting two or more photon pairs within a given timing window. These multiple photon pairs are uncorrelated since the characteristic timing window of our detection system ($t_{c}=$2.4 ns) is much larger than the coherence time of the photons ($<$1 ps). Once the source pair rate R becomes large compared to the timing window ( $R\times t_{c} > 0.1$), the multi-pairs lead to a dominant amount of accidental coincidences, thereby creating a noise floor which covers the entangled correlations and reduces the observed entanglement visibility. This limitation is solely due to the emitted pair number, almost irrespective of attenuation. With our source we observe that, given the 2.4 ns coincidence window implemented in our analyzing detection electronics, the observable entanglement visibility therefore drops below 80\% at pump power of merely 2.2 mW (generating about 50 Mpairs/second). Higher pair rates are possible using higher pump powers, but the accidental coincidences would reduce the entanglement visibility below the threshold required by applications (typically 70\% cutoff). Even for optimistically short coincidence windows of 100 ps\cite{Gallivanoni:06}, a visibility of merely 80\% would be achievable with our compact source pumped at 40 mW (see Fig. \ref{multipairs}). This clearly demonstrates that significant advancements in terms of useful entangled photon pair rates require further improvements in the field of detector technology and timing electronics.

\section{Conclusion}
By careful design of the focusing and spectral characteristics of the pump beam and the SPDC photons originating from two crossed PPKTP crystals, we achieve a detected pair rate of 640 kcp/mW, a detected spectral brightness of 278 kcp/mW/nm and a Bell state fidelity of 0.98. Combined with the compact footprint and with the simplicity and ruggedness of the configuration, these results make the source an ideal device for future field experiments on quantum communications and quantum entanglement tests, and well-suited for mobile and space applications. The benefit of using one of the highest nonlinear coefficients via noncritical quasi phase-matching in KTP, as well as the collinear emission geometry providing the optimal configuration for high collection efficiencies from long crystals, enable us to reach for the maximal brightness expected in schemes based on bulk SPDC. Furthermore, we showed that the potential pair creation rate of our source leads to a photon flux which extends beyond the capabilities of current single-photon detectors and commercially available timing electronics.

\section*{Acknowledgements}

We thank Josep M. Perdigues and Eric Wille of the European Space Agency for valuable discussions. Project funding EQUO ESTEC Contract N.: AO/1-5942/08/NL/EM and contracts TEC 2010-14832 and FIS2011-23520 funded by the Spanish Ministry of Science and Innovation. HW, RU and JR acknowledge funding by the EU-project Q-ESSENCE. JPT acknowledges support from the Government of Spain (FIS2010-14831) and the project PHORBITECH (FET-Open grant number 255914).

\bibliographystyle{IEEEtran}

\end{document}